\documentclass[manuscript]{acmart}

\usepackage{CJKutf8}

\AtBeginDocument{%
  }

\setcopyright{acmlicensed}
\copyrightyear{2018}
\acmYear{2018}
\acmDOI{XXXXXXX.XXXXXXX}
\acmConference[Conference acronym 'XX]{Make sure to enter the correct
  conference title from your rights confirmation email}{June 03--05,
  2018}{Woodstock, NY}
\acmISBN{978-1-4503-XXXX-X/2018/06}

\begin{document}

\title{\textit{Vocabuild}: An Accessible Augmented Tangible Interface for Gamified Vocabulary Learning of Constructing Meaning}
\renewcommand{\shorttitle}{Vocabuild}

\author{Siying Hu}
\affiliation{%
  \department{Department of Computer Science}
  \institution{City University of Hong Kong}
  \city{Hong Kong SAR}
  \country{Hong Kong SAR}
}

\author{Zhenhao Zhang}
\affiliation{%
  \department{Department of Computer Science}
  \institution{City University of Hong Kong}
  \city{Hong Kong SAR}
  \country{Hong Kong SAR}
}

\renewcommand{\shortauthors}{Hu and Zhang}

\begin{abstract}
Vocabulary acquisition in early education often relies on rote memorization and passive screen-based tools, which can fail to engage students kinesthetically and collaboratively. This paper introduces \textit{Vocabuild}, an augmented tangible interface designed to transform vocabulary learning into an active, embodied, and playful experience. The system combines physical letter blocks with a projection-augmented surface. As children physically construct words with the blocks, the system provides real-time, dynamic feedback, such as displaying corresponding images and animations, thus helping them construct semantic meaning. Deployed in a classroom context, our gamified approach fosters both individual exploration and peer collaboration. A user study conducted with elementary school children demonstrates that our tangible interface leads to higher engagement, increased collaboration, and a more positive attitude towards learning compared to traditional methods. Our contributions are twofold: (1) the design and implementation of \textit{Vocabuild}, a projection-augmented tangible system that transforms vocabulary learning into an embodied and collaborative activity; and (2) empirical findings from a classroom study showing that our tangible approach significantly increases engagement, peer collaboration, and positive learning attitudes compared to traditional methods.

\end{abstract}

\ccsdesc[500]{Human-centered computing~Interactive systems and tools}

\keywords{Tangible User Interface; Language Learning; Accessibility; Deaf and Hard-of-Hearing; Embodied Cognition; Gamification; RFID}

\begin{teaserfigure}
  \includegraphics[width=\textwidth]{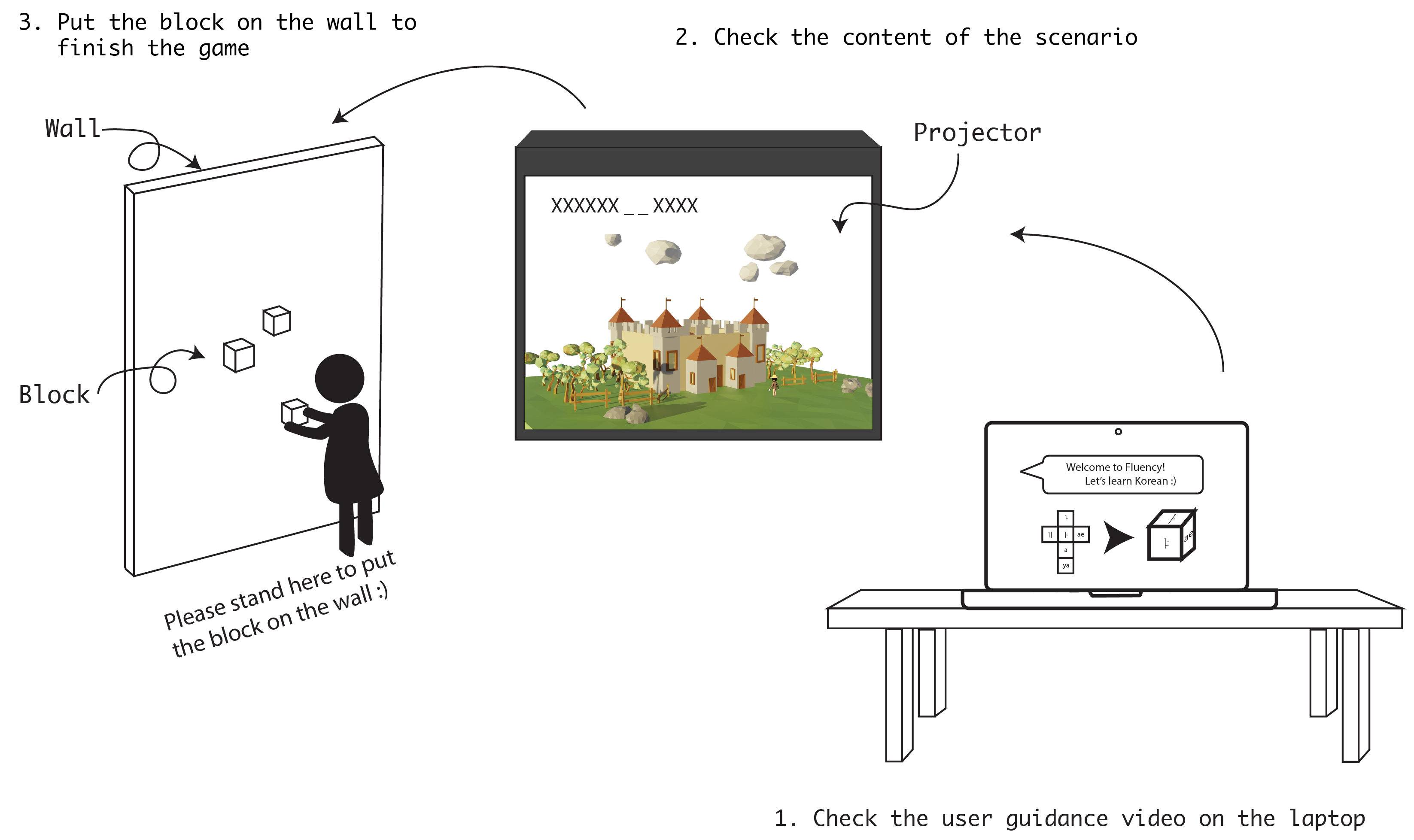}
  \caption{Children's vocabulary learning should be more than passive screen tapping. We introduce \textit{Vocabuild}, a tangible augmented interface that combines physical letter blocks with dynamic visual feedback. By physically building words, children transform the abstract task of spelling into a concrete, playful, and creative game.}
  \label{fig:teaser}
\end{teaserfigure}

\received{20 February 2007}
\received[revised]{12 March 2009}
\received[accepted]{5 June 2009}

\maketitle

\section{Introduction}
Early language acquisition is a cornerstone of human development, foundational to cognitive growth, social integration, and future academic success \cite{weisleder2013talking}. For most children, this journey begins as an immersive auditory experience. They learn by listening, associating the sounds of spoken words with objects, actions, and emotions. Later, this auditory foundation becomes critical for literacy, as children develop phonological awareness, the ability to recognize and work with sounds in spoken language, which is a strong predictor of reading ability \cite{lonigan2000emergent, ehri2001meta}.

However, as learning increasingly moves to digital platforms, a closer examination reveals a critical disconnect. Many popular language-learning applications, while promising educational enrichment, often merely digitize traditional, passive learning methods. They reduce language to a series of abstract drills on a flat screen—tapping flashcards, tracing letters, and completing repetitive matching exercises \cite{hirshpasek2015apps, meyer2021educational}. This approach fundamentally clashes with the established principles of early childhood pedagogy, which emphasize that young children learn best through physical manipulation, sensory exploration, and active play \cite{fan2024multisensory, lockman2021interactions}. By confining learning to a disembodied, two-dimensional space, these tools risk transforming what should be a joyful process of discovery into a chore of passively receiving and memorizing abstract symbols, rather than an active game of creating tangible meaning.

To address this critical gap, we turn to the fields of Tangible User Interfaces (TUIs) and Embodied Cognition. TUIs bridge the digital and physical worlds by allowing users to interact with digital information through tangible objects \cite{liang2025value}. This approach aligns powerfully with the theory of Embodied Cognition, which posits that learning and cognition are not abstract mental processes but are deeply rooted in the physical interactions of the body with the environment \cite{faella2025educational}. By physically manipulating objects that represent abstract concepts, learners can form stronger, more intuitive mental models \cite{gilliganlee2023hands}. We argue that this tangible, embodied approach holds immense potential for creating a more equitable and effective learning pathway for children, allowing them to "construct meaning" through a modality in which they are fluent: touch and physical action.

This paper introduces \textit{Vocabuild}, an accessible tangible interface for gamified vocabulary learning, designed to meet the developmental needs of kids learners. \textit{Vocabuild} consists of a set of physical blocks embedded with RFID tags, a reader, and a large-screen display powered by a game engine. The core interaction is inherently visual-first and non-auditory. Children learn by physically selecting and arranging blocks, each representing a character, to construct a word. The system provides immediate, rich, and context-aware visual feedback on the screen, creating a complete learning loop that does not depend on sound. Inspired by the ancient invention of movable type printing, \textit{Vocabuild} transforms the abstract task of spelling into a concrete, creative act of building.

Through the design, development, and preliminary evaluation of \textit{Vocabuild} in a simulated classroom environment, we make a three-fold contribution to the HCI community:

\begin{itemize}
\item \textbf{A System Contribution:} We present the complete design and implementation of \textit{Vocabuild}, a tangible learning system that offers a replicable, accessible alternative to conventional screen-based language tools.
\item \textbf{A Design Contribution:} We provide a case study and design implications for creating tangible learning tools for young children, demonstrating how a `visual-first, non-auditory' feedback model can effectively support vocabulary learning.
\item \textbf{An Empirical Contribution:} We offer an empirical example of how Embodied Cognition can be leveraged to create more inclusive educational technologies. We show how mapping the physical act of constructing a word to its visual meaning can lower the barrier to entry for learners of diverse abilities, thereby expanding the application of TUIs in early childhood education.
\end{itemize}

\section{Related Work}

\paragraph{Digital Language Learning for Young Children}

The proliferation of screen-based devices has led to a surge in digital language learning applications \cite{duolingo2024whitepaper, gilligan2023appbased, alqahtani2024evaluation}. Yet, many designed for young children fail to align with their developmental needs, often digitizing traditional, passive learning methods \cite{AHMAD2021CRI, hirshpasek2015apps}. This approach reduces language acquisition to a series of disembodied interactions like tapping flashcards or tracing letters on a flat screen \cite{sari2023flashcard, chu2024digitalplay}. It positions the child as a passive recipient of information rather than an active constructor of knowledge, clashing with established pedagogical principles that emphasize the importance of physical manipulation, sensory exploration, and play in early childhood education \cite{liang2025value}. Furthermore, many of these tools are heavily reliant on auditory feedback, creating an accessibility barrier for Deaf and Hard-of-Hearing (DHH) learners \cite{alit2025systematic} and potentially inducing a split-attention effect that increases cognitive load for all children \cite{10.1007/s11042-022-12296-2}. This reveals a critical gap for a learning tool that is embodied, accessible, and empowers children as active agents in their learning process.

\paragraph{Tangible Interfaces for Learning and Embodied Cognition}

Tangible User Interfaces (TUIs) offer a powerful paradigm to bridge this gap by coupling digital information with physical objects, allowing users to ``think with their hands'' \cite{mustar2025tui}. This approach is strongly supported by the theory of Embodied Cognition, which posits that cognitive processes are deeply rooted in the body's physical interactions with the environment \cite{faella2025educational}. By enabling learners to manipulate physical representations of abstract concepts, TUIs can lower cognitive barriers and help form stronger, more intuitive mental models \cite{gordon2021embodied}. Previous research has explored TUIs for language learning, such as using physical blocks for spelling \cite{10.1145/2317956.2318016, liu2025tares}. However, \textit{Vocabuild} extends this work in two critical dimensions. First, it introduces a generative and narrative feedback model. Instead of providing simple binary feedback (correct/incorrect), the system responds to mistakes that form other valid words (e.g., spelling water instead of horse) with the corresponding meaningful animation. This transforms errors into "unexpected discoveries," fostering resilience and deeper semantic links. Second, \textit{Vocabuild} is explicitly designed around a visual-first, non-auditory principle, making it inherently accessible and demonstrating a robust design pattern for inclusive learning technologies that has been underexplored in tangible language tools.

\paragraph{Constructionism and Learning Through Play}

Our design is grounded in Seymour Papert's educational philosophy of Constructionism \cite{papert1980mindstorms, papert1991situating, ackermann2001constructionism}, which argues that learning occurs most effectively when individuals are engaged in constructing a personally meaningful, shareable artifact \cite{10.1145/3715336.3735737}. \textit{Vocabuild} embodies this by transforming the task of spelling into an act of building a word. The physical word and its resulting on-screen animation become the tangible artifact, shifting the child's role from a student completing an exercise to a creator authoring meaning. This sense of agency and creation is a powerful intrinsic motivator \cite{mwinsa2025playbased, resnick2017lifelong}. The experience is framed as Learning Through Play, which is the primary and most natural way children explore ideas and interact with their world \cite{frost2001play}. By focusing on the joy of discovery and the immediate, rewarding feedback of seeing their creations come to life, our system turns vocabulary acquisition from a rote memorization chore into an intrinsically motivating game of exploration and discovery, fostering a positive and engaged attitude toward learning.

\section{System Design and Implementation}
To instantiate our vision of a tangible, playful, and accessible language learning experience, we designed and implemented \textit{Vocabuild}. This section details the design goals that guided our process, the technical architecture of the system, and the step-by-step interaction flow for the user. Our design process was iterative, moving between low-fidelity prototypes and functional implementations to refine both the physical and digital aspects of the experience.

\subsection{Design Goals}
Our design was driven by three core principles derived from our review of educational theory and existing learning tools.

\subsubsection{Accessibility First}
A primary goal was to create an experience that was not dependent on auditory feedback, making it inherently inclusive. This ``visual-first, non-auditory'' aims ensures that the entire learning loop—from action to feedback, which can be completed using only visual and tactile information. This approach benefits not only learners who may have hearing impairments but also reduces the cognitive load for all young children by providing clear, unambiguous feedback in a single sensory channel. Furthermore, accessibility was considered in the physical form factor, such as the blocks were designed to be large and lightweight, with clear, high-contrast characters, making them easy to handle for children with developing fine motor skills.

\subsubsection{Embodied Interaction}
This goal drawing from the theory of Embodied Cognition \cite{shapiro2019embodied, sullivan2018learning}, we aimed to map the abstract task of spelling onto a concrete physical action. The goal was to allow children to think with their hands. Instead of tapping on a screen, the learner engages in the physical act of grasping, orienting, and arranging blocks. This process leverages kinesthetic memory and spatial reasoning to reinforce the sequence of characters in a word. The act of physically building a word is intended to create a stronger cognitive link to its meaning than simply viewing it on a screen.

\subsubsection{Engaging Experience}
To transform learning from a chore into play, our third goal was to foster deep engagement through principles of game design and constructionism \cite{kafai2012constructionism, kafai2006playing}. The experience was designed to provide intrinsic motivation through the joy of creation—the user is an active builder, not a passive recipient. This is supplemented by extrinsic motivation in the form of clear goals (e.g., helping an on-screen character achieve a task) and immediate, rewarding feedback. The aesthetic design, featuring colorful blocks and dynamic animations, was carefully crafted to spark curiosity and create a sense of wonder, encouraging sustained exploration and experimentation.

\subsection{System Architecture}
\textit{Vocabuild}'s architecture integrates custom-built physical components with sensing hardware and a responsive digital environment. The system is designed to be a standalone unit suitable for a classroom setting.

\subsubsection{Physical Prototype Fabrication} 
\begin{figure}
    \centering
    \includegraphics[width=1\linewidth]{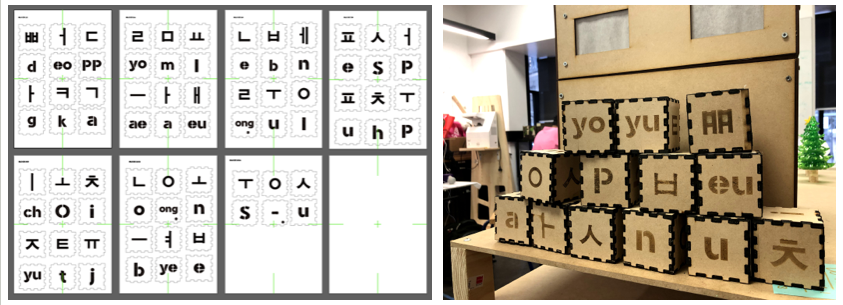}
    \caption{
    The physical interaction components of \textit{Vocabuild}. The blocks are laser-cut, designed to be hollow and lightweight for easy handling by children. Each block features a double-sided design: one side is engraved with a Korean character (e.g., ``\begin{CJK}{UTF8}{mj}커\end{CJK}''), while the other shows its Romanized phonetic equivalent (e.g., ``keo''). This design aims to help children bridge the grapheme-phoneme correspondence in early learning stages and supports varied learning strategies.
    }
    \label{fig:physical}
\end{figure}

We used laser-cut MDF for its light weight and RFID signal penetration. Each block housed a 13.56MHz RFID tag and was hollow to be lightweight and easy for children to handle. The reading station was constructed as a custom shelf or frame, designed to hold the blocks in a linear sequence and discreetly house the sensing electronics beneath the surface (can see in \autoref{fig:physical}).

\subsubsection{Sensing and Electronics}
The core of the sensing system is an MFRC522 RFID reader module connected to a Particle Photon microcontroller via the SPI (Serial Peripheral Interface) protocol. 
A Particle Photon microcontroller, chosen for its Wi-Fi and Arduino compatibility, managed an RFID reader via SPI. The entire electronics assembly was powered by a portable power bank, creating a self-contained and wire-free interactive unit that could be easily deployed in various settings.

\subsubsection{Digital Prototype Development}

\begin{figure}
    \centering
    \includegraphics[width=1\linewidth]{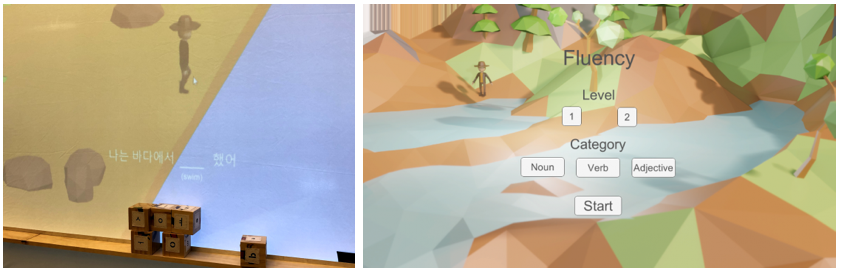}
    \caption{The \textit{Vocabuild} system in use and its digital user interface. (Left) The digital prototype developed in Unity, featuring a gamified learning scenario and a real-time display of the digital characters corresponding to the physical blocks. (Right) A child interacts with the system by arranging the physical blocks on the reading station to complete a word-building task. The system directly maps the child's physical actions to dynamic visual outcomes on the screen, creating a complete and non-auditory embodied learning loop.}
    \label{fig:digital}
\end{figure}

The system software consists of two parts. First, microcontroller firmware written in C++/Wiring continuously scans for RFID tags, concatenates their UIDs, and sends them to a host computer via USB serial. Second, a C\# application in Unity listens to the serial port, parses the UIDs, and uses a dictionary to map the UID sequence to a corresponding word. A state machine then triggers the appropriate visual feedback, such as animations and UI updates (can see in \autoref{fig:digital}).

\subsection{Interaction Flow}
The interaction flow begins with the system presenting a gamified challenge on the screen, which typically involves a visual scenario with a corresponding sentence that has a single word missing. To guide the learner, a `visual cheat sheet' is placed near the interface, allowing users to reference the characters needed for the target word. The user then selects the physical blocks and physically constructs the word by placing them in sequence on the reading station. As the blocks are arranged, the system provides real-time, one-to-one mapping on the screen, displaying the digital character for each physical block placed. Once the user finalizes their construction, the system evaluates the word and delivers the appropriate visual feedback from the three-tiered model. This cycle of physical action and immediate, nuanced feedback allows the learner to actively test hypotheses, understand the outcomes of their constructions, and iterate, transforming the abstract task of spelling into a concrete and engaging process of discovery.

\section{User Study}

We conducted a qualitative field study in a simulated classroom to evaluate \textit{Vocabuild}’s usability and learning value for young children follow prior work \cite{wolins1992children, fry1997qualitative}. Our study addressed three research questions:

\begin{itemize}
    \item How do children utilize the tangible interaction to understand vocabulary meaning?
    \item How do children utilize the tangible interaction to construct vocabulary meaning?
    \item How do teachers perceive the system's role and potential within a classroom teaching dynamic?
\end{itemize}

\subsection{Participants}

We recruited three teachers and five children (3 male, 2 female; aged 5-7). The teachers each had over three years of professional experience in early childhood education or language instruction. All children were native English speakers with no prior Korean instruction. Before the study, we obtained informed consent from teachers and parents, and verbal assent from each child.

\subsection{Procedure and Task}
\begin{CJK}{UTF8}{mj}
Each study session, which involved one teacher and one or two children, was designed to last approximately 30 minutes. To serve as a visual scaffold for the children, a printed poster, referred to as the visual cheat sheet, was placed next to the apparatus, showing the target words alongside images of the corresponding character blocks. The procedure was methodically structured into three distinct phases to observe different aspects of interaction. The session began with a five-minute introduction, during which the teacher framed the \textit{Vocabuild} system as a block game and explained that the goal was to help an on-screen character by building words. This was followed by a 5 minute period of free exploration, where children were encouraged to freely handle and arrange the blocks without a specific task, allowing us to observe their initial curiosity and exploratory patterns. The main part of the session was a 10 minute guided task completion phase, where the teacher led the child through two or three word-building challenges, such as spelling ``커피'' (coffee). The teacher provided hints and directed the child's attention to the visual cheat sheet as needed. 
\end{CJK}

\subsection{Data Collection and Analysis}
We video-recorded all sessions and took observational field notes. After each session, we conducted a 30-minute semi-structured interview with the teacher and student. Two researchers then performed a thematic analysis on the triangulated data (videos, notes, and interview transcripts) to identify recurring themes \cite{trombeta2022textual}.

\section{Results}
Through a thematic analysis of our video recordings, observational notes, kids and teacher interviews, we identified core themes that illuminate the \textit{Vocabuild} system's effectiveness and user experience. These findings address our research questions regarding engagement, the process of meaning construction, and the evolving role of the teacher in a technology-augmented classroom.

\subsection{Fostering Engagement and Exploration through Physical Manipulation}

Our primary observation was that the physical, tangible nature of the \textit{Vocabuild} system served as a powerful catalyst for immediate and sustained engagement. Unlike typical screen-based applications, the blocks invited physical exploration and transformed the learning task into an act of play.

\subsubsection{Transforming Passive Observation into Active Exploration}
All five children were immediately drawn to the physical blocks, often before paying close attention to the on-screen prompt. Notably, this interaction was entirely intuitive. Without any external prompting or guidance, the children spontaneously began to assemble the blocks, pausing to inspect the letters and patterns on their surfaces even if they did not recognize them. This deep physical engagement appeared to significantly prolong their attention spans, creating a focused and self-directed interaction loop. One of the teachers highlighted this stark contrast with digital-only tools \textit{``These days, we use a lot of tablets for teaching, and you'll see kids swiping and tapping away. But after a few minutes, their eyes usually glaze over, maybe bored or tired of the same old routine. But I saw this one kid, he spent nearly twenty minutes straight playing with this wooden block, just to make sure the order was right. Totally focused, not distracted by anything else around him, or losing interest when he made a mistake and moving on.''}.

\subsubsection{Intrinsic Motivation through Playful Discovery}
The system successfully framed the learning activity as a game rather than a chore. The children perceived the blocks as toys with a power to change the screen and play the game, which intrinsically motivated them to experiment. This experimentation manifested as a deliberate physical loop in which children would use their eyes and hands to search for a printed block, grasp it to consider its meaning or category, and then place it according to their own intentions. A 6-year-old participant articulated this sense of empowerment and discovery: \textit{``Chose this, then that, and then (...) poof match. The pics and characters on the screen changed, not sure if it's right, but hey, at least I can keep playing till I get it right!''}. This cycle of seeking and placing concentrated their attention and effectively lowered the anxiety associated with learning a new language.

\subsection{Constructing Meaning Through Action and Feedback}
The system’s core interaction loop enabled children to actively construct meaning of  vocabulary. We identified four strategies through which children used trial-and-error to build an understanding of the relationship between characters, words, and their semantic outcomes.

\subsubsection{Strategy 1: Scaffolding through Direct Visual Matching}

We observed all children initially used direct visual matching, methodically finding physical blocks that matched the characters on the visual cheat sheet to build the target word. This process was repeated for each character in sequence.
This strategy had the immediate effect of demystifying the task and building foundational confidence. It allowed the child to achieve a successful outcome quickly, providing a crucial dose of positive reinforcement that motivated them to continue interacting. It effectively served as a built-in tutorial, teaching the core mechanics of the system without explicit instruction.

\subsubsection{Strategy 2: Learning Syntax through Sequential Self-Correction}
\begin{CJK}{UTF8}{mj}
When a child made a mistake in the sequence, the real-time on-screen display of the spelled characters provided immediate and granular feedback. For instance, while trying to spell ``커피'' (coffee), a child might correctly place ``커'' but then incorrectly place ``스''. They would look up at the screen and see the wrong character appear. This discrepancy prompted a focused, micro-correction. Instead of clearing all the blocks and starting over, the child would correctly identify the single point of failure, remove only the incorrect ``스'' block, and resume their search for the correct ``피`` block. The child learned that not only does each block have a meaning, but its position in the sequence is critical. The ability to isolate and fix a single error without penalty reinforced a one-to-one mapping between the physical block and its digital representation, and it made the process of debugging a word feel manageable and non-punitive. It transformed a potential moment of frustration into a small, solvable puzzle. A 7-year-old verbalized this self-correction process perfectly while fixing a word: \textit{``The screen shows a circle on top, but my block made a line. Okay, so not this one (...) I see it on the screen now, it matches.''}
\end{CJK}

\subsubsection{Strategy 3: Deepening Semantic Links through Trial-and-Error Discovery}
\begin{CJK}{UTF8}{mj}
The most profound learning moments occurred when a child's spelling error resulted in a different, but still valid, word. For example, when the target was ``말'' (mal, horse), a child accidentally spelled ``물'' (mul, water) due to the visual similarity of the characters. It's a word with an up-down structure. Their expectation was to see a horse animation. Instead, the system triggered a fully realized animation of the on-screen character drinking a glass of water. This created a moment of cognitive dissonance, a violation of their expectation that prompted immediate reflection. A teacher described the unique pedagogical power of this feedback:
\begin{quote}
    \textit{``This trial-and-error approach is pretty cool (...) it shows positive results instead of error messages that might discourage kids. Even if they spell something wrong, they still get to see the animation... seeing the character actually drink water(...) is a little learning moment. Maybe they’ll remember this word better.”''}.
\end{quote}
This non-punitive, consequential feedback was exceptionally effective at creating strong and memorable semantic links. The abstract spelling error was translated into a concrete, narrative event. The child didn't just learn that ``물'' was the wrong spelling for ``horse'', they learned that ``물'' means water. This transformed a mistake from a failure into an unexpected discovery, fostering intellectual curiosity and resilience. It taught them that the system of language has its own internal logic, where different combinations yield different, valid meanings.
\end{CJK}

\subsubsection{Shifting from Following Rules to Creating Language}
\begin{CJK}{UTF8}{mj}
After several successful attempts, we observed a clear behavioral shift as children moved away from relying on the cheat sheet. They began to internalize the block-character associations and developed an emerging kinesthetic memory, often picking up a block and knowing its character without needing to inspect it. This newfound fluency paved the way for a more advanced strategy which we identified as hypothesis testing. Children would intentionally combine familiar blocks in novel sequences, not to complete a given task, but to proactively discover what would happen. This experimental play marked their crucial transition from being passive learners to active creators. They were no longer just decoding instructions but were beginning to encode their own ideas, a process that fostered a powerful sense of agency and authorship. For instance, when a child successfully created a new word not on the cheat sheet, like forming ``코'' (nose), they experienced the thrill of linguistic creation.
\end{CJK}

\subsection{Redefining the Teacher's Role through AI Collaboration}
The \textit{Vocabuild} system fundamentally shifted the teacher's role within the learning environment, transforming them from a primary instructor into a facilitator and co-explorer.

\subsubsection{Offloading Repetitive Labor to Enable High-Level Focus}
Our interviews revealed that the system significantly reduced the teachers' workload by automating the most repetitive and time-consuming aspects of language instruction. Teachers noted that a large portion of their effort in traditional drills was spent on monotonous, low-level tasks, such as presenting problems, verifying answers, and providing repetitive corrective feedback. This process was described as draining and inefficient. The \textit{Vocabuild} system effectively took over this role, acting as a tireless assistant that could be delegated the responsibility for foundational instruction, such as rote memorization of vocabulary. This liberation from monotonous loops freed up significant time and cognitive energy, allowing teachers to shift their focus towards higher-order pedagogical goals. As one teacher articulated:\textit{``My typical role in a language drill is being a human feedback machine, Yes, that's right, No, try again. It can take over me. I can see someone need help.''}, this change was profound.

\subsubsection{Enabling Targeted Scaffolding for Deeper Understanding}
With the system handling the core mechanics of feedback and repetition, teachers were able to dedicate their full attention to diagnosing the unique learning needs of each child. Instead of being caught in a cycle of correcting surface-level errors like spelling, they could observe why a child was struggling and provide highly targeted, personalized guidance. For example, a teacher could design impromptu activities that helped a student better understand the contextual use of a word, rather than just its spelling. As one teacher explained:\textit{``I rarely get a chance to step back and take a good look at the whole class, analyzing what they all have in common when it comes to learning. Most of the time, it's feedback that makes me spend extra time thinking, which might help me spot issues some kids are having right away.''}. This allowed them to act as a precise educational partner, intervening at the most critical moments to foster deeper conceptual understanding.

\subsubsection{Envisioning a Future of Embodied and Collaborative Vocabulary Teaching}
Teachers envisioned a more profound collaborative future, where the system could evolve from a tool into a proactive learning agent. They speculated about assigning the AI agent to introduce foundational concepts to students before a formal lesson. They suggested assigning the AI to introduce foundational concepts as pre-class work, allowing students to arrive already prepared for higher-level activities. This shift would transform valuable class time from rote instruction into a dynamic workshop for creative application and problem-solving. One teacher summarized this potential:\textit{``I could assign this as homework so they learn the basics before class. Our time together then becomes about creating, not just repeating. It would let me guide their curiosity, not just check their answers.''}. The teacher's role would thus be elevated from an instructor to a facilitator of deeper learning.

\section{Discussion}
\textit{Vocabuild} reshapes learning by grounding abstract spelling in physical manipulation, fostering an engaged and exploratory environment.

\subsection{Rethinking Roles in the Tangible Learning Environment}
Our study highlights a fundamental shift in the role of educational technology, moving from a simple instructional tool to a dynamic partner for both students and teachers. For students, \textit{Vocabuild} functioned not as a rigid tutor, but as a embodied play partner. The tangible blocks and immediate feedback loop empowered them with a sense of agency, transforming them from passive rule-followers into active creators who could form and test their own hypotheses, a principle central to constructionist learning theories \cite{holman1997rethinking, kynigos2015constructionism}. This shift is crucial for fostering intrinsic motivation and intellectual curiosity \cite{talib2009instructional, oudeyer2016intrinsic}. However, this partnership is not without risks. An over-reliance on direct scaffolding could potentially encourage surface-level matching rather than deeper semantic understanding, a known challenge in designing learning supports \cite{thorpe2002rethinking, so2010designing, gan2023large}. The system's design should consider carefully balance guidance with challenges that promote genuine cognitive effort.

From the teachers' perspective, the system evolved into a valuable collaborator. By automating the repetitive labor of drills and corrections, \textit{Vocabuild} offloaded the monotonous aspects of instruction, acting as a tireless assistant. This liberated teachers to focus on higher-order pedagogical tasks, which means diagnosing individual learning challenges, providing targeted scaffolding \cite{kress2014teaching}, and facilitating creative exploration \cite{samson2015fostering, ferrari2009innovation}. This elevates the teacher's role from a sage on the stage to a guide on the side, a long-standing goal in educational reform \cite{samson2015fostering, so2010designing}. If teachers become too removed from the foundational learning process, they might lose insight into students' core difficulties, a concern raised in literature on human-AI collaboration \cite{han2025helping}. The ideal dynamic is a true partnership, where the technology handles the mechanics, enabling the teacher to provide nuanced, human-centered guidance.

\subsection{Design Implications for Embodied and Accessible Education}

Beyond the system itself, our work present the visual-first and non-auditory feedback model as a robust design pattern. This approach is inherently inclusive, directly benefiting learners who are Deaf or Hard-of-Hearing by aligning with visual learning strengths \cite{marschark2011evidence, cheng2024motion}. More broadly, it reduces the cognitive load for young children by providing clear, unambiguous feedback within a single sensory channel. This avoids the split-attention effect that can occur when learners must process conflicting or redundant information from different modalities \cite{yoon2011effects}. The instance where a child immediately self-corrected a misplaced block based solely on the on-screen visual discrepancy provides strong evidence for the efficacy of this model.

We provides a compelling empirical example of applying Embodied Cognition theory to lower cognitive barriers in language learning same as prior work \cite{lan2015embodied, atkinson2010extended, damiani2017embodied}. The physical act of grasping, orienting, and sequencing blocks transforms the abstract challenge of spelling into a concrete, manageable task, a core principle of tangible interaction \cite{10.1145/2317956.2318016, liu2025tares, 10.1145/3170427.3188576}. This process leverages kinesthetic memory and spatial reasoning to reinforce word structures. The mistake of spelling water instead of horse illustrates this. The system's response—a meaningful animation for the wrong word—turned a potential moment of failure into an unexpected discovery. This non-punitive, consequential feedback creates a powerful and memorable link between a physical action and its semantic outcome, demonstrating how tangible interaction can make abstract concepts accessible and learning resilient for users of all abilities.

\subsection{Limitations}
Our study has several limitations that offer avenues for future research. The primary limitation is the small sample size of five children and three teachers, which means our qualitative findings on engagement and learning strategies are insightful but not statistically generalizable. The study was also short-term, with each session lasting only twenty minutes in a simulated classroom. This makes it difficult to assess long-term learning retention or to rule out the potential influence of the novelty effect on student engagement. Finally, our evaluation focused on beginner-level Korean vocabulary, and further research is needed to understand how the \textit{Vocabuild} system scales to more complex linguistic concepts and adapts to the dynamics of a real, less-controlled classroom environment.

\section{Conclusion and Future Work}
This paper introduced \textit{Vocabuild}, a tangible augmented interface that transforms vocabulary acquisition into an embodied, playful, and collaborative activity. Our study demonstrates that by mapping the physical act of building words to immediate visual feedback, the system significantly enhances engagement and fosters a positive attitude towards learning, allowing children to transition from passive observers to active creators who construct meaning through exploration and self-correction. Building on these promising results, we plan to conduct a long-term, in-situ deployment to evaluate learning retention and classroom integration. Future work will also focus on expanding the system's capabilities to include more complex grammar and adaptive challenges, evolving it into a personalized learning agent that collaborates with teachers. By continuing to bridge the physical and digital worlds, we aim to create more inclusive and effective educational tools that empower all learners.

\begin{acks}
To Robert, for the bagels and explaining CMYK and color spaces.
\end{acks}

\bibliographystyle{ACM-Reference-Format}
\bibliography{Main}


\end{document}